\documentclass[useAMS,usenatbib]{mn2e}

\usepackage{times}
\usepackage{graphicx}

\newcommand{\beq}{\begin{equation}}
\newcommand{\eeq}{\end{equation}}

\newcommand{\om}{\omega}

\newcommand{\bfep}{\mbox{\boldmath $\epsilon$}}
\newcommand{\bfgam}{\mbox{\boldmath $\gamma$}}

\newcommand{\bfg}{\mathbf{g}}

\newcommand{\1}{\Omega_{\rm M} }
\newcommand{\2}{\Omega_{\rm v} }
\newcommand{\4}{\Omega_{\Lambda} }

\def\gs{\mathrel{\lower0.6ex\hbox{$\buildrel {\textstyle >}\over{\scriptstyle \sim}$}}}
\def\ls{\mathrel{\lower0.6ex\hbox{$\buildrel {\textstyle <}\over{\scriptstyle \sim}$}}}

\begin{document}

\title[Triplet statistics]{Measuring dark energy with the shear triplet statistics}
\author[M. Sereno]{M. Sereno\thanks{E-mail:sereno@physik.unizh.ch} 
\\
Institut f\"{u}r Theoretische Physik, Universit\"{a}t Z\"{u}rich,
Winterthurerstrasse 190, CH-8057 Z\"{u}rich , Switzerland
}

\date{February 1, 2006}

\maketitle

\begin{abstract}
The shear triplet statistics is a geometric method to measure cosmological parameters with observations in the weak gravitational lensing regime towards massive haloes. Here, this proposal is considered to probe the dark energy equation of state and its time derivative in view of future wide-field galaxy surveys. A survey with a median redshift of $\sim 0.7$ and a total area of $\sim 10000$ square degrees would be pretty effective in determining the dark matter cosmological density and in putting useful constraints on the dark energy properties.
\end{abstract}


\section{Introduction}

The deep theoretical understanding of gravitational lensing makes it an attractive probe of dark energy, one of the main puzzle of modern cosmology. Dark energy can show up either in a pure geometrical way by affecting the distance redshift relation or via its effect on the growth of structure. Lensing is sensitive to both signatures since a ratio of distances appears in the scaling of lensing parameters with redshifts and because the effective mass of lensing structures reflects the power spectrum and growth rate of large-scale density perturbations \cite[and references therein]{ba+sc01,mun+al06}. Probing dark energy through both geometry and growth by taking weak gravitational lensing two-point functions (such as the shear power spectrum) of distant galaxy images as a function of redshift can be appealing \cite[and references therein]{hea+al06,hut02,so+kn04,ta+ja04}, but this method relies on the interpretation of the distortion signal on scales where non linear evolution necessitates accurate modeling from numerical simulations. In the strong lensing regime the distance ratio and the cosmological parameters can be derived using observations of giant luminous arcs in rich clusters of galaxies but the mass profile and the mass normalization of the lenses must be assumed to be known, for example from X-ray measurements \cite[and references therein]{ser02,se+lo04}. Strong gravitational lensing observations of image separations and time delays could provide useful constraints as well \citep{lin04}.

The basic idea behind geometric methods is to separate out the information purely from the distance ratios, irrespective of the lensing mass distribution. In the strong lensing regime, this can be achieved by considering multiple images systems and the ratios of arc angular positions, but this still requires the assumption that the lensing mass, even if unknown, is pretty regular \citep{gol+al02}. The weak lensing regime seems more promising. \citet{ja+ta03} proposed to take the ratio of the galaxy-shear correlation function at different redshifts behind galaxy groups and galaxy clusters. Dark energy parameter constraints based on this geometric method, which extends into the nonlinear matter power spectrum but still drops out the mass of the lens, have been then discussed by several authors in view of future galaxy surveys \citep{be+ja04,hu+ja04,so+kn04}. The method was then further developed by directly considering ratios of the shears behind a cluster without first generating the cross-correlation functions \citep{tay+al07}. Alternatively, \citet{zha+al05} pointed out how the cross-correlation of a foreground galaxy-density field or shear field with the shear field from a background source population scales with the source redshift in a way that can be used to constrain cosmology without making assumptions about the mass/galaxy power spectrum. 

In this paper, we want to reconsider the so called triplet statistics, an original idea to constrain cosmological parameters from weak lensing in galaxy clusters proposed in \citet{gau+al00}. This method is able to disentangle the effect of the lensing mass, described by local convergence and shear terms, from the cosmological parameters by considering the ellipticities in triplets of galaxies located at about the same angular position but having different redshifts. Differently from similar proposals \citep{tay+al07}, the triplet statistics is not limited to the outskirts of massive haloes and can be used even in the inner regions. \cite{gau+al00} originally considered the method for determining the cosmological constant with observations towards very massive galaxy clusters. Here, we review the triplet statistics and discuss the measurement of the dark energy properties with future weak lensing survey.

\section{Basics}
\label{sec:basi}

The distortion of images of background galaxies is determined by the convergence $k$, i.e. the lensing strength, and the complex shear $\bfgam = \gamma_1 + i \gamma_2$. The lensing parameters can be related to the value they would have for a source at a reference redshift \citep{se+sc97,gau+al00},
\begin{eqnarray}
k     &  = & \om (z_\mathrm{s}) k_\mathrm{ref} ,       \label{basi1}  \\
\bfgam & = & \om (z_\mathrm{s}) \bfgam_\mathrm{ref}  . \label{basi2}
\end{eqnarray}
The lensing factor $\om (z_\mathrm{s})$ is defined as a ratio of angular diameter distances and contains the cosmological dependency,
\beq
\label{basi3}
\om (z_\mathrm{s}) \equiv  \frac{D_\mathrm{ds} }{ D_\mathrm{os} } \left(   \left. \frac{D_\mathrm{ds} }{D_\mathrm{os}} \right|_{z_\mathrm{s} = z_\mathrm{ref}}   \right)^{-1},
\eeq
where $D_{ij}$ is the angular diameter distance between the redshifts $z_i$ and $z_j$ with the redshifts of interest being those of the observer $o$, the lens $d$ and the source $s$.  We consider a standard Friedmann-Lema\^{\i}tre-Robertson-Walker model of universe, filled with non-interacting pressure-less matter and dark energy, parameterized through its equation of state $w \simeq w_o+w_a(1-a)$, with $a \equiv 1/(1+z)$ . Since the contribution from relativistic particles is negligible in the redshift range investigated in our analysis, we will neglect it in what follows. In such a model of universe, the angular diameter distance between an observer at $z_i$ and a source at $z_j$ is
\beq
\label{basi4}
D_{ij}=\frac{c}{H_0}\frac{1}{1+z_j}\frac{1}{|\Omega_{\rm K0}|^{1/2}} {\rm Sinn} \left( \int_{z_i}^{z_j} \frac{|\Omega_{\rm K0}|^{1/2}}{H (z)/H_0} dz \right),
\eeq
where
\begin{equation}
\frac{H(z)}{H_0}= \sqrt{ 
\1 a^{-3}+ \2 a^{-3(1+w_0+w_a)} e^{-3 w_a (1-a)} +\Omega_{\rm K} a^{-2} },\nonumber
\end{equation}
and $H_0$ is the present value of the  Hubble parameter; $\1$ and $\2$ are the today normalized densities of dust and dark energy, respectively; $\Omega_{\rm K} \equiv 1- \1 -\2$; Sinn is defined as being sinh when $\Omega_{\rm K}>0$, sin when $\Omega_{\rm K}<0$, and as the identity when $\Omega_{\rm K}=0$. For the expression of the distance in an inhomogeneous universe, we refer to \citet{ser+al01,ser+al02}.

The transformation from intrinsic to observed ellipticity in the weak lensing regime ($\gamma , k \ll 1$) takes a simple form \citep{se+sc97,gau+al00}. Due to a lensing halo, a galaxy with intrinsic ellipticity $\bfep_\mathrm{s}$ is imaged with ellipticity \citep{gau+al00}
\beq
\label{basi5}
\bfep \simeq (1-g^2)\bfep_\mathrm{s} +\bfg
\eeq
with $\bfg \equiv \bfgam /(1-k)$ being the complex reduced shear. Background galaxies at (nearly) the same angular position probe the same local cluster mass distribution and gravitational potential. Then comparing the shear amplitude for three galaxies having different redshift allows to separate the effect of the mass distribution from cosmology. A geometrical operator can be built from the measured ellipticities $\bfep_i$ and redshifts $z_i$ of the three galaxies $i=\left\{ a,b,c \right\}$ in such a way that it depends only on cosmology \citep{gau+al00}
\beq
\label{basi6}
T_{abc} = \left|
\begin{array}{ccc}
1 & \om_a & \om_a \bfep_b \bfep_c^* \\
1 & \om_b & \om_b \bfep_c \bfep_a^* \\
1 & \om_c & \om_c \bfep_a \bfep_b^* 
\end{array}
\right| .
\eeq
Redshifts of the galaxies inside triplets are in ascending order. For three intrinsically spherical galaxies, $T_{abc} = 0$ when the lensing factors $\omega_i$ are computed for the actual values of the cosmological parameters, apart from noise. $T_{abc}$ is linear regarding to the ellipticities, which makes the principal source of noise randomly distributed around zero. Apart from noise, the main part of the triplet operator contains the cosmological dependence and can be approximated as
\beq
\label{basi7}
T^\mathrm{m}_{abc} \simeq \left|
\begin{array}{ccc}
1 & \om_a & \om_a/\om_a^0 \\
1 & \om_b & \om_b/\om_b^0 \\
1 & \om_c & \om_c/\om_c^0 
\end{array}
\right| \om_a^0 \om_b^0 \om_c^0 \gamma_\mathrm{ref}^2 ,
\eeq
where the apex $0$ denotes that the angular diameter distances have been calculated for the actual values of the cosmological parameters and where we have considered only one component of the shear. If we neglect the contribution from the local convergence, the reduced sher $\bfg$ can be identified with the local shear and then it is enough to take the ratio of the observed ellipticities of a pair of near galaxies in order to separate the effect of cosmology \citep{tay+al07}. Neglecting $k$ introduces a systematic error that can be significant for the largest clusters. A $10\%$ variation of the cosmological parameters changes the reduced shear by about $1\%$ which is ten times smaller than the relative variation due to the $1-\om k_\mathrm{ref}$ for $k_\mathrm{ref} \ls 0.1$ \citep{gau+al00}. Even if the majority of the signal comes from intermediate mass clusters, the highest return in accuracy comes from the largest haloes \citep{tay+al07} so that this effect must be properly accounted for.

Let us sort the redshift in a triplet such that $z_a < z_b < z_c$. From the matrix form of the operator $T_{abc}$, it is clear that if two galaxies in a triplet are very close, then $T_{abc} \simeq 0$ with no regard to the cosmological parameters. $T_{abc} \simeq 0$ also if the minimum redshift in the triplet is very close to the lens redshift. Once fixed the minimum and the maximum redshift in a triplet, the sensitivity of the operator $T_{abc}$ is maximized for an intermediate redshift $z_b$ nearly in the middle of the redshift range. Once fixed $z_a$ and $z_b$, the sensitivity increases with the maximum redshift $z_c$. Hence, the main information from the triplet method comes from the high redshift tail of the background source distribution.

\section{Forecast for lensing surveys}
\label{sec:fore}

\begin{figure}
        \resizebox{\hsize}{!}{\includegraphics{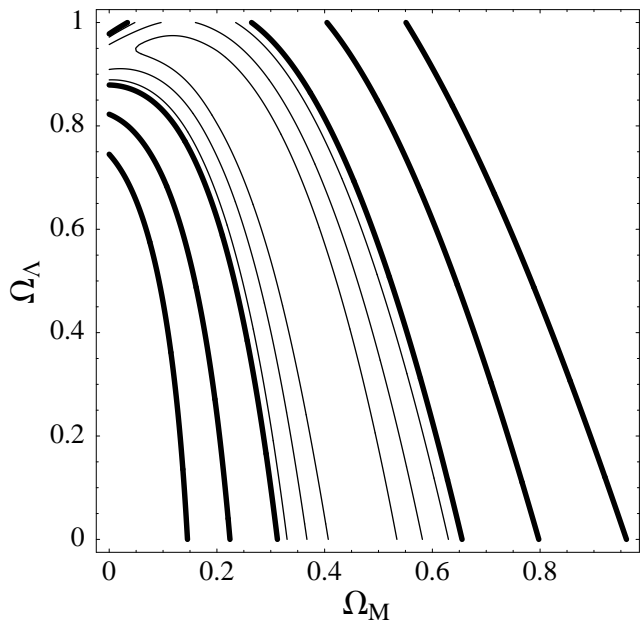}}
        \caption{Confidence regions in the $\1-\4$ plane as expected for a lensing survey with $z_\mathrm{m} \sim 0.7$. Contours show the 1,2-3$\sigma$ limits around the fiducial model $\1=0.3, \4=0.7$. The inner thin and the outer thick contours refer to a survey with total area $A=10000~\mathrm{deg}^2$ and $1500~\mathrm{deg}^2$, respectively.}
        \label{om_ol}
\end{figure}

\begin{figure}
        \resizebox{\hsize}{!}{\includegraphics{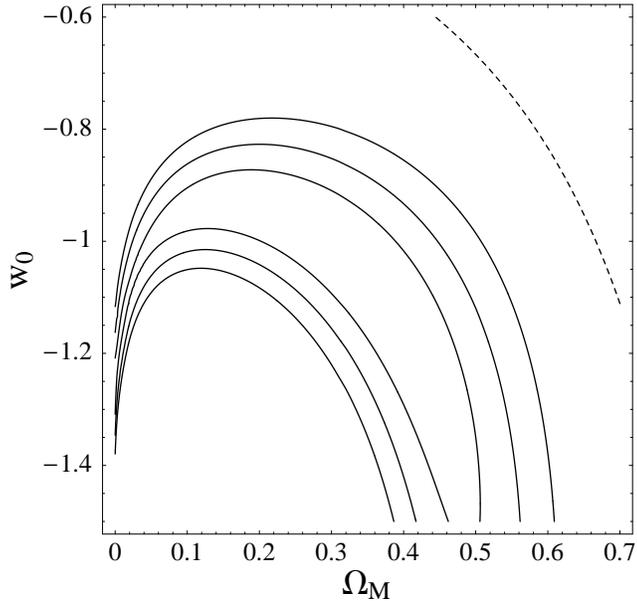}}
        \caption{Confidence regions in the $\1-w_0$ plane for a lensing survey covering $10000~\mathrm{deg}^2$. Contours show the 1,2-3$\sigma$ limits around the fiducial flat model $\1=0, w_0=-1$ and assuming no evolution for the dark energy, $w_a=0$. The dashed line separate models with either accelerated or decelerated expansion.}
        \label{om_w0}
\end{figure}

\begin{figure}
        \resizebox{\hsize}{!}{\includegraphics{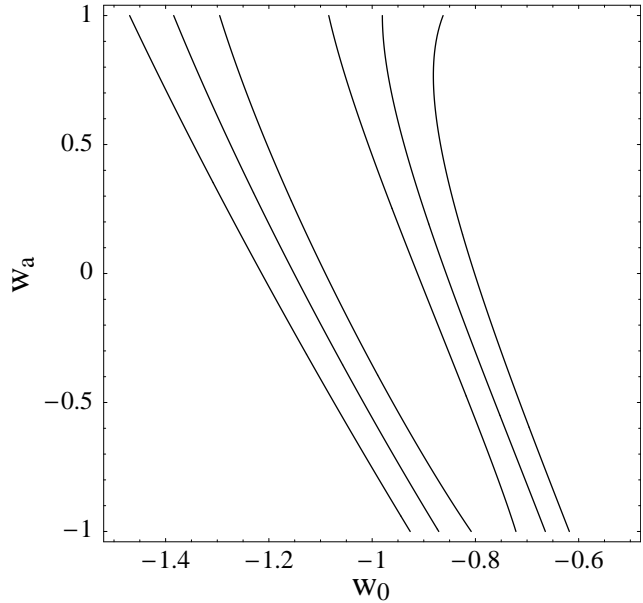}}
        \caption{Confidence regions in the $w_0-w_a$ plane for a lensing survey covering $10000~\mathrm{deg}^2$. Contours show the 1,2-3$\sigma$ limits around a reference flat model $w_0=-1,w_a=0$ assuming flat geometry and a sharp prior $\1=0.3$. }
        \label{w0_wa}
\end{figure}

There are a number of current and planned imaging surveys for weak lensing analyses \citep{pea+al06}. Beyond 2007, the funded VST (VLT Survey telescope) public survey KIDS will cover at least 1500 square degrees in 4 broadband optical filters. Combined with a following near infrared coverage by VISTA, this will yield a 9-band optical-IR survey with depth approximately 2 mag deeper than the Sloan and with very accurate photometric redshift estimate. The typical KIDS lenses will be at $z \sim 0.2-0.4$. Beyond KIDS, the planned Dark Energy Survey on the CTIO Blanco telescope or the darkCAM survey at VISTA are expected to take another steps forwards in terms of sky coverage, imaging $\sim 10000$ square degrees. In what follows, we give a cosmological parameter estimation forecast for such surveys.

The background redshift distribution for a typical magnitude-limited survey can be taken to be
\beq
\label{fore1}
\frac{dn_\mathrm{gal}}{dz} = n_0 \frac{3}{2} \frac{z^2}{z_0^3} \exp \left[ - \left( \frac{z}{z_0} \right)^{3/2} \right] .
\eeq
The redshift scale $z_0$ is related to the median redshift of the survey, $z_\mathrm{m}$, by $z_0=z_\mathrm{m}/1.412$; $n_0$ gives the total number density of sources with usable photometric redshift and shape estimate. For a KIDS-like survey we can take $z_\mathrm{m} \sim 0.7$.

The main part of the lensing signal comes from halo lensing masses in the range $5\times 10^{13} \ls M/M_\odot \ls 10^{15}$. The number density of haloes can be accurately calculated and varies from $n_\mathrm{hal} \gs 10^2$ per square degree for $M \sim 10^{13}$ to $n_\mathrm{hal} \gs 10^{-2}$ per square degree for $M \sim 10^{15}$ \citep{tay+al07}. An aperture size $\theta_\mathrm{max} \simeq 2'-5'$ corresponds roughly to the virial radii of such massive haloes over the redshift range considered. An inner circle with aperture $\theta_\mathrm{in} \simeq 0.1'-0.8'$ has to be excised from the data to exclude the arclets and the strong lensing regime. Considering a simple isothermal sphere model for the lenses, we get that $\langle \gamma_\mathrm{ref} \rangle \sim 0.04 - 0.15$ in the mass range of interest. We remark as the triplet operator is proportional to $\langle \gamma_\mathrm{ref}^2 \rangle$ rather than $\langle \gamma_\mathrm{ref} \rangle^2$. These estimates for the shear are lower limits since considering a Navarro-Frank-White profile as a deflector, the shear signal would increases for the same halo mass.

In order to study the sensitivity of the triplet method on the cosmological parameters, we perform a $\chi^2$ analysis. The main difference with the statistical analysis in \cite{gau+al00} is that they considered a mean triplet operator as the average over all triplets whereas here we consider a $\chi^2$ built on the linearly independent triplets. Not all of the triplets we can put together behind a lens contain independent information. It can be easily shown that from $N_g$ near galaxies only $N_g-2$ triplets out of the binomial factor $\left( N_g, 3 \right)$ are independent. If a triplet contains at least one galaxy not already included in the sample of other triplets, then it is linearly independent. Triplet selection can be properly optimized. As maximum distance between the triplet components we take a separation of $\Delta r \sim 20''$ \citep{gau+al00}. This is a good balance between having a sufficient number of triplets and not smearing the signal. As $\chi^2$, we consider
\beq
\label{fore2}
\chi^2 = \sum_{l,\{ a,b,c \} } \left( \frac{T_{abc}^\mathrm{m}}{\delta T_{abc}} \right)^2.
\eeq
with the sum running over the lensing haloes $l$ optically selected in the survey and the independent triplets for each halo. Foreground lenses and background sources are modeled according to the previous discussion. For each background galaxy we completed the triplet by selecting the two neighboring galaxies ($\Delta r \ls 20''$) which maximize the signal. 

Statistical and systematics errors affecting the method have been deeply discussed in \cite{gau+al00}. The main sources of noise are: $i)$ the intrinsic source ellipticities; $ii)$ the errors on measured ellipticities; $iii)$ the fact that sources do not experience exactly the same potential and finally $iv)$ the errors on measured (photometric) redshifts. It can be shown that due to the linearity of the operator, the noise is linear with respect to each individual term and then is proportional to $1/\sqrt{N}$ with $N$ the total number of triplets. The dominant contribution to the error budget comes from the intrinsic ellipticity. 

Together with statistical noise, several systematics affect the method. The main ones are well understood and have been identified as $i)$ a bias due to an asymmetry in the probability distribution of the terms $\Delta \om$ due to photometric redshift errors and, mainly, $ii)$ contamination by background structure, either galaxy-galaxy lensing or large-scale structure. Sources in very close angular pairs, for which another galaxy could play the role of lens, could be rejected  from the sample. It can be shown that for surveys large enough the effect of large-scale structure can be neglected at first order with respect to the statistical noise \citep{gau+al00,tay+al07}. The dominant noise source is by due to intrinsic ellipticity dispersion \citep{gau+al00,tay+al07}. Then, as an estimate of $\delta T_{abc}$ in Eq.~(\ref{fore2}) we will take
\beq
\label{fore3}
\delta T_{abc}^2 = \frac{\sigma_\epsilon^2}{2} \sum_{i=a,b,c} \left. \left( \frac{\partial T_{abc}  }{\partial \epsilon_i} \right)^2 \right|_{ \epsilon_i = \omega_i^0 \gamma_\mathrm{ref}},
\eeq
where the factor 2 in the denominator arises because we are using only one component of the measured ellipticity. As intrinsic ellipticity dispersion we take $\sigma_\epsilon =0.3$. The effect of lensing by large-scale structure will be discussed later. However, even if we are underestimating the statistical noise by considering only the main contributor, the choice of parameters has been generally conservative. Results for a survey with median redshift $\sim 0.7$ and a typical density of $ n_0 \sim 30$ galaxies per square arcminute are shown in Figs.~\ref{om_ol},~\ref{om_w0} and \ref{w0_wa}. As a first step we have considered dark energy in the form of a cosmological constant ($w_0=-1$ and $w_a=0$), see Fig.~\ref{om_ol}. The contours show the 1, 2 and 3$\sigma$ confidence limits for two parameters ($\Delta \chi^2=2.30,6.16$ and $11.8$ respectively). With a survey area of $10000$ square degrees, the triplet statistics can constrain the matter density parameter but is pretty insensitive to the total amount of vacuum energy. This degeneracy changes with redshift. For tests in the strong lensing regime based on luminous giant arcs, the contours are nearly orthogonal to the triplet estimator since the distance ratio is evaluated for very high redshift sources \citep{se+lo04}.

Assuming a flat model and a constant equation of state, the triplet statistics can put a quite firm upper limit on $w_0$, see Fig.~\ref{om_w0}. On the other hand, the confidence regions spreads out well within the phantom regime ($w_0 < -1$), as usual for methods based on the distance-redshift relation. The phantom regime would be also consistent with an higher value of the matter density parameter. Contours in the $w_0-w_a$ plane are still pretty elongated, see Fig.~\ref{w0_wa}. On its own, the triplet statistics can not say much about the evolution with time of the dark energy, but if combined with orthogonal methods such as the cosmic microwave background radiation or the baryonic acoustic oscillation in the matter power spectrum, constraints could be significant.

The linear size of the confidence regions shrinks approximately as $1/\sqrt{N_g}$, with $N_g$ the total number of galaxies in the survey. Cosmological constraints then would strongly benefit by a very large survey area. Furthermore, increasing the median redshift of the survey would both increase the local galaxy density and would probe the distance ratio in a redshift range more sensitive to cosmology. 
 
Together with the variance term proportional to the intrinsic uncertainty per shear mode due to the galaxy intrinsic ellipticities ($\sigma_\gamma$), which is related to the shape noise and to the shot noise, the other main source of error is due to lensing by large-scale structure in between the lens and the sources ($\sigma_\mathrm{LLS}$), related to the sampling variance term \citep{zha+al05,hu+ja04}. Depending on the survey strategy, these two terms can be comparable. \citet{tay+al07} gave a simple approximate scaling relation between the two terms for a survey of median redshift $z_\mathrm{m}$, $\sigma_\mathrm{LSS}^2 \simeq ( 24.1 z_\mathrm{m}^4 \Delta z ) \sigma_\gamma^2 $, where $\Delta z$ is the typical photometric redshift error for the survey. Then, for $z_\mathrm{m} \sim 0.7$ and $\Delta z \sim 0.05$, error estimates on cosmological parameters obtained considering just the shot noise should be increased by $\sim 15\%$.

\section{Final remarks}

The aim of this paper has been the evaluation of the triplet statistics as a dark energy probe in view of future galaxy survey. The main source of statistical uncertainty in our statistical approach was the intrinsic source ellipticity. As regards the main systematics, the contribution of large-scale structure to the observed shear, being uncorrelated with the cluster effect, should average out over independent clusters along different lines of sight \citep{tay+al07}. In a survey measuring photometric redshifts, data should be collected in redshift bins with width equal to the typical redshift error at that redshift. Then, the number of independent triplets behind a cluster would be $N_{bin}-2$, with $N_{bin}$ the total number of bins. However, photometric redshifts, especially if some infrared filters are available, should be accurately determined. 

In this paper, we have considered any halo mass profile but the method could be optimized using the symmetrical properties in the mass distribution of galaxy groups and clusters. A nearly elliptical matter distribution would allow to consider tangential shear averaged in concentric annuli, i.e. to collect triplets selecting galaxies in the concentric ring instead of a small local patch. This would make nearly sure that for any galaxy we can find a pair of galaxy redshifts that maximize the signal.

Some conclusions on the viability of the shear triplet method can be drawn by comparison with the shear ratio geometric test. The confidence regions we plotted seem larger that those obtained with the shear ratio test in a survey with similar properties \citep{tay+al07}. As we have seen the shear test is biased for large mass haloes $M \gs 10^{15} M_\odot$, where the reduced shear should be properly considered instead of the shear. Being these haloes pretty rare if compared with halo masses of order of $\ls 10^{14} M_\odot$, which provide the bulk of the signal, this systematic effect should not jeopardize the shear ratio method. In any case, since the two techniques require the same kind of measurements they should be properly integrated. This is desirable especially because the analysis of optically selected large mass halo in wide field survey should begin with very massive haloes which, on turn, pay the highest dividend.

\section*{Acknowledgments}
M.S. is supported by the Swiss National Science Foundation and by the Tomalla Foundation.

%\bibliographystyle{mn2e}
%\bibliography{triplets}

\setlength{\bibhang}{2.0em}

\end{document}